\documentclass[aps,prb,twocolumn,superscriptaddress,showpacs,floatfix]{revtex4}
\usepackage{comment}
\usepackage{graphicx}
\usepackage{latexsym}
\usepackage{stmaryrd}
\usepackage{amsmath}
\usepackage{amssymb}
\usepackage{epstopdf}
\usepackage{bm}
\usepackage[utf8]{inputenc}
\usepackage[colorlinks,linkcolor=red,anchorcolor=red,citecolor=blue]{hyperref}
\usepackage{url}

\begin{document}

\title{Topological Layer-Spin Filter in Screw Dislocation}

\author{Jiaojiao Zhou}
\affiliation{School of Mathematics and Physics, Anhui Jianzhu University, Hefei 230601, China}
\author{Hong Hu}
\affiliation{School of Mathematics and Physics, Anhui Jianzhu University, Hefei 230601, China}
\author{Jiangying Yu}
\affiliation{School of Mathematics and Physics, Anhui Jianzhu University, Hefei 230601, China}
\author{Lin Xu}
\affiliation{Center of Free Electron Laser $\&$ High Magnetic Field, Information Materials and Intelligent Sensing Laboratory of Anhui Province, Institutes of Physical Science and Information Technology, Anhui University, Hefei 230601, China}
\affiliation{Leibniz International Joint Research Center of Materials Sciences of Anhui Province, Hefei 230601, China}
\author{Shu-guang Cheng}
 \email{sgcheng@nwu.edu.cn}
\affiliation{Department of Physics, Northwest University, Xi'an 710069, China}
\author{Hua Jiang}
 \email{jianghuaphy@fudan.edu.cn}
\affiliation{Interdisciplinary Center for Theoretical Physics and Information Sciences (ICTPIS), Fudan University, Shanghai 200433, China}
\date{\today}

\begin{abstract}
While the quantum spin Hall effect leverages two-dimensional topological states to manipulate spin without dissipation, layertonics extends this paradigm to three dimension by enabling control over the layer degree of freedom. Topological materials incorporating screw dislocations exhibit the capability for simultaneous manipulation of both electronic spin and layer degrees of freedom. In this work, the electronic transport properties of a multilayer Kane-Mele model with screw dislocations is studied theoretically. Numerical simulations of a screw dislocation reveal that dissipationless quantum spin Hall edge states propagate not only at the outer boundaries of the structure but also along the screw dislocation itself, working as layer-spin filter. In detail, 1) the spin-up and spin-down carriers starting from the same source layer flow to different drain layers along the topological channels, respectively. 2) The spin of carriers flowing into a given drain layer is determined by the input source layer. Moreover, we found that the transmission coefficient and spin polarization remain robust against Anderson disorder. Under magnetic disorder, spin flip and backscattering occur, suppressing the transmission coefficient while maintaining nearly unchanged spin polarization. Finally, the layer- and spin-resolved transport properties in a device with two screw dislocations are investigated as well. We have developed an innovative methodology to modulate electron transport with simultaneous layer and spin resolution.
\end{abstract}

\maketitle
\section{INTRODUCTION}
Numerous unique topological states have been identified in graphene materials \cite{CNAH,KM,DJR,DPU,NKF,BLF}, including quantum spin Hall (QSH) edge states. The QSH effect facilitates the transport of momentum-spin locked currents through helical edge states \cite{KM,BBAZ,MJHTL,YAF,WCH,BJK}. At a single sample boundary, spin-up and spin-down currents propagate in opposite directions, exhibiting robustness against non-magnetic disorders\cite{XuMoore,ShindouMurakami}. In multilayer graphene systems, the structural complexity and interlayer interactions have led to increasingly diverse research on quantum states\cite{ZQ,IM,NJC,XCBL,JuL,LiS,CSG,YCVA,GCL,YCVS,GCA}.
These states are not merely superpositions of the quantum states of individual layers; instead, interlayer interactions drive the emergence of novel quantum phase.

Beyond planar stacking, graphene can form unique structural configurations, one of which is the screw dislocation, a distinctive crystal defect in multilayer graphene\cite{HGR,PAR,WZJ}. In this configuration, multilayer graphene forms a helicoidal surface, accompanied by a continuous spiral structure that extends through it. Moreover, the Burgers vector, characterizing the interlayer spacing in multilayer graphene, is much larger than the carbon-carbon bond length\cite{HGR, NIP}. Consequently, both interlayer van der Waals interactions and electron-electron interactions are substantially weaker than the coplanar carbon-carbon covalent bonding.
Therefore, electrons can flow on this seamless and three-dimensional network, leading to exotic transport properties.
Generalizing from graphene systems, screw dislocations are commonly observed in various two-dimensional materials\cite{ZYZ,ZYZ2,ZYZ3,CLL} and three-dimensional topological insulators\cite{AKN,HH,LYMW,CCZH,LLS,CYTY}.
Recent studies have shown that screw dislocations can influence the electrical properties of materials by altering their local electronic structure. For example, abrupt variations in the local density of states\cite{AKN} and conductivity\cite{HH} have been observed at screw dislocations, indicating the presence of one-dimensional propagating states. Similar propagation modes have also been demonstrated in classical waveguide systems, including photonic crystals \cite{LE,LFF} and phononic crystals\cite{YLQC,WHCY,ZYD,LZK}.

Significant efforts have been made in experiment to eliminate the screw dislocations in multilayer materials for the purpose of high quality samples\cite{add1,add2,add3,add4,add5}. However, given the topological nature of many two-dimensional materials, the intentional incorporation of such screw dislocations could lead to exotic spin transport properties.
Given the three-dimensional network nature of the system, charge separation with layer degrees of freedom becomes achievable, establishing the foundation for layertronics\cite{AGao,DZhai,WBDai,SLiMGong}: the controlled manipulation of layer-indexed quantum states.
Finally, the interplay between layer and spin degrees of freedom may enable highly tunable layer- and spin-resolved electron transport, offering exciting opportunities for exploring novel physical effects.

In this work, we first investigate this phenomenon using the Kane-Mele model in graphene with a screw dislocation. We find that the topological channels are located at the outer boundaries of multilayer graphene and along the screw dislocation, enabling the QSH edge states to flow to adjacent layers or transport between the top and bottom layers. In this multilayer structure, carriers with different spins will be transmitted to different drain layers. This implements a layer-spin filter. This mechanism is supported by the numerical calculations. Second, we examine the effect of disorder on the transport properties of these channels. Furthermore, we extended this layer-spin selection mechanism to systems containing two screw dislocations. This work demonstrates the potential application of screw dislocations in multilayer graphene-like materials to enable layer-resolved spin filtering, offering new insights into topological quantum states and their applications in spintronics and layertronics.

This paper is organized as follows. Section II introduces the Kane-Mele model with a screw dislocation and the numerical methods for calculating (spin) linear conductance, spin polarization, and spatial current distribution. The key spin and layer transport characteristics for systems with single and double screw dislocations are presented in Sec. III. Finally, a brief conclusion is presented in Sec. IV.
\begin{figure}[b]
\includegraphics[scale=0.195]{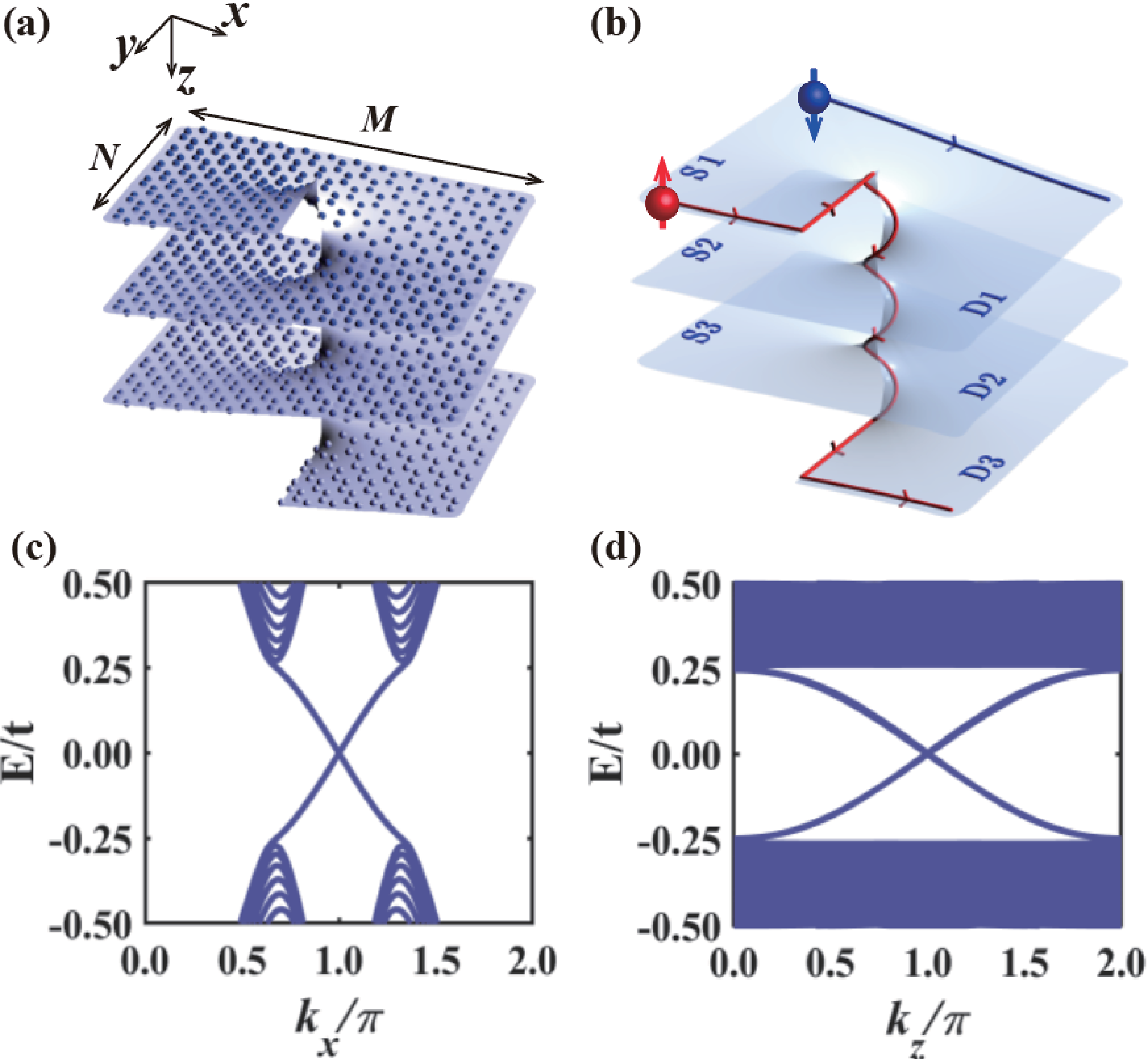}
\caption{\label{fig:1}
(a) Schematic of the three-layer graphene based structure with the screw dislocation located at the center of the \(x\)-\(y\) plane. The lattices are represented by the blue balls. The sample sizes are represented by $M=13$ along the $x$ direction, and $N=6$ along the $y$ direction.
(b) The topological channels of QSH edge states. The blue line indicates the outer topological edge state, while the red line corresponds to the topological state located at the screw dislocation. The source terminals (S1, S2, S3) and drain terminals (D1, D2, D3) are labeled here.
(c) The band structure of the Kane-Mele nanoribbon along the $x$-axis. It corresponds to the QSH edge states localized at the outer boundary of the structure in (b).
(d) The band structure of the screw dislocation [see the red line in (b) for the QSH edge states]. The state at the outer boundary is eliminated by applying the periodic boundary conditions in the $x$- and $y$-directions.}
\end{figure}

\section{MODEL AND METHODS}

The platform of the layer-spin filter we adopt is a three-layer graphene structure with a screw dislocation, as shown in Fig.~\ref{fig:1}(a).
Here, only the bonding effects of covalent bonds are considered, i.e., the inter-layer coupling is ignored.
Besides the central region shown in Fig. \ref{fig:1}(a), the structure also includes six terminals.
For each layer there are two terminals along the \(x\)-direction.
The sources on the left side are labeled S1, S2, and S3 sequentially for different layers, while the drain layers are labeled D1, D2, and D3 accordingly.
The tight-binding Hamiltonian of the Kane-Mele model is described by \cite{KM}:

\begin{eqnarray}\label{eq1}
&&H=H_0+H_{AD/MD},\nonumber\\
&&H_0=t\sum_{\langle \bf{i}, \bf{j} \rangle}c_{\bf{i}}^+\sigma_0c_{\bf{j}}+i\lambda_{SO}\sum_{\langle\langle \bf{i},\bf{j} \rangle \rangle}\upsilon_{\bf{ij}}c_{\bf{i}}^+\sigma_zc_{\bf{j}}+h.c., \nonumber\\
&&H_{AD}=\sum_{\bf{i}}w_{\bf{i}}c_{\bf{i}}^\dagger \sigma_0c_{\bf{i}}, \nonumber\\
&&H_{MD}=\sum_{\bf{i}}Wc_{\bf{i}}^\dagger {\bf{S_n}} c_{\bf{i}}.
\end{eqnarray}

The total Hamiltonian comprises two components: the clean part $H_0$ and the disordered part, which includes Anderson disorder $H_{AD}$ or magnetic disorder $H_{MD}$.
In $H_0$, $\langle \bf{i},\bf{j}\rangle$ and $\langle\langle \bf{i},\bf{j}\rangle \rangle$ describe the nearest and the next-nearest-neighbor hopping from site $\bf{j}$ to site $\bf{i}$. The hopping energy $t$ is adopted as the unit energy and $\sigma$ is the Pauli matrix.
The first term of $H_0$ describes the nearest coupling of a pristine graphene. The second term of $H_0$ is the spin-orbit coupling strength $\lambda_{SO}$ where $\lambda_{SO}=0.05t$ is used throughout this work. $\upsilon_{\bf{ij}}=1(-1)$ is the Haldane factor for clockwise(counterclockwise) next-nearest-neighbor coupling\cite{Hf}.
In the disorder part, $w_{\bf{i}}$ is randomly distributed in $[-W/2,W/2]$ at site $\bf{i}$ where $W$ is the disorder strength. For $H_{MD}$, $\bf{S_n}$ is given by $\bf{S_n}=\bf{\sigma}\cdot\bf{n}$ with the unit vector $\bf{n}$ representing the magnetic orientation at site $\bf{i}$. Here, the polar angle $\theta_{\bf{i}}$ is randomly distributed in $[0,\pi]$ and the azimuthal angle $\phi_{\bf{i}}$ is randomly distributed in $[0,2\pi)$.

To characterize the transmission coefficient for electrons transmitting from source layer $p$ with spin-$\alpha$ to drain layer $q$ with spin-$\beta$, the zero-temperature spin conductance $G^{\alpha\beta}_{pq}(E)$ at the Fermi energy $E$ is calculated. Based on the non-equilibrium Green??s function method and the Landauer-B{\"u}ttiker formula \cite{Datta}, the zero-temperature spin-resolved linear conductance under an infinitesimal bias can be expressed as
\begin{equation}
G^{\alpha\beta}_{pq}(E)=\frac{e^2}{h} \mathrm{Tr}[{\bf{\Gamma}}_{p\alpha}(E){\bf G}^r(E){\bf{\Gamma}}_{q\beta}(E){\bf G}^a(E)].
\end{equation}
Here, $e$ is the elementary charge of an electron, and $h$ is Planck's constant. ${\bf G}^r$/${\bf G}^a$ is the retarded/advanced Green's function, and is calculated from the relation ${\bf G}^r=[{\bf G}^a]^\dagger= (E{\bf{I}}-{\bf{H}}_c-{\bf{\Sigma}}^r_p-{\bf{\Sigma}}^r_q)^{-1}$. ${\bf{H}}_c$ is the Hamiltonian of the central region and $\bf{\Sigma^r_{p/q}}$ is the retarded self-energy of source/drain terminals\cite{lee,sancho}. To calculate the linear conductance for a specific spin, the source and drain materials are assumed to be pristine graphene nanoribbons in which the spin index is well defined. Thus, the line-width functions for a given spin index ($\bf{\Gamma_{p\alpha}}$ and $\bf{\Gamma_{q\alpha}}$ defined from the self-energy), are calculated.
The total conductance from source layer $p$ to drain layer $q$ is
\begin{equation}
G_{pq}(E)=\sum\limits_{\alpha,\beta} G_{pq}^{\alpha\beta}(E).
\end{equation}

The spin polarization of transmission coefficients from source layer $p$ to the drain layer $q$ is defined as
\begin{equation}\label{P}
  P_{pq}=\left | \frac{G_{pq}^{\downarrow\uparrow}+G_{pq}^{\uparrow\uparrow}-G_{pq}^{\downarrow\downarrow}-G_{pq}^{\uparrow\downarrow}} {G_{pq}}\right |.
\end{equation}

The local current from the site $\boldsymbol {i}$ to site $\boldsymbol {j}$ can be expressed as \cite{x1,x2,cheng2}.
\begin{equation}\label{current}
J^\alpha_{\boldsymbol {i}\boldsymbol {j}}(E)=\frac{2e^2V}{h}\mathrm{Im}\{\sum_{\beta}H_{\boldsymbol {i}\alpha, \boldsymbol {j}\beta}[{\bf G}^r(E)\Gamma_{p\alpha}(E){\bf G}^a(E)]_{\boldsymbol {j}\beta,\boldsymbol {i}\alpha}\}.
\end{equation}
Here $H_{\boldsymbol {i}\alpha,\boldsymbol {j}\beta}$ is the coupling between neighbouring sites ${\bf{i}}$ and ${\bf{j}}$, including the nearest-coupling term and the next-nearest-coupling term. The total current is proportional to the small voltage bias $V=V_p-V_q$ between the source and drain terminals. Moreover, along any line that cuts across the nanoribbon, the sum of the local currents should be conserved.

\section{RESULTS AND DISCUSSION}

Before delving into the transport properties of layer-spin filter, we first present the energy bands of two systems based on the Kane-Mele model. They are a single-layer graphene nanoribbon and a multilayer graphene with a screw dislocation, as shown in Fig. \ref{fig:1}(c) and \ref{fig:1}(d). For the first case, the width of nanoribbon is $N=25$. For the later case, periodic condition is assumed in both the $x$ and $y$ directions to outline the role of screw dislocation. The energy band of both structures exhibit a topological gap in the range of $[-3\sqrt{3}\lambda_{SO},3\sqrt{3}\lambda_{SO}]$, which corresponds to $[-0.26t,0.26t]$ when $\lambda_{SO}=0.05t$.
In details, for the graphene nanoribbon, if the Fermi energy $E$ lies within this gap, there are a pair of QSH edge states propagating along the outer boundaries [see Fig. \ref{fig:1}(c)]\cite{KM}. In contrast, for the model in Fig. \ref{fig:1}(d), the topological states travels along the screw dislocation. The one-dimensional gapless state along the screw dislocation provide a clear explanation for the abrupt changes in conductance and local density of states observed in previous experimental studies\cite{HH,AKN}.

For a finite-sized three-layer graphene structure with a screw dislocation, the topological channels exist both at the outer boundaries of the structure and at the screw dislocation, as shown in Fig.~\ref{fig:1}(b). Consequently, spin polarized current can flow along the outer boundary (marked in blue) to adjacent layers or transfer from the top layer to the bottom layer via the screw dislocation (marked in red). The mechanism of the layer-spin filter in Fig.~\ref{fig:1}(b) operates as follows: the drain layers D1, D2, D3 are all attached and the input layer can be either S1, S2 or S3. Taking S1 as the input source as an example, one branch (marked by the blue line) carry spin-down state propagates from S1 to D1 via the back edge of the top layer. The other branch (marked by the red line) carrying spin-up state, travels from the front edge of the top layer to the bottom layer via the screw dislocation located at the center of device. Thus, these topological channels serve as the foundation for realizing a topological layer-spin filter.

\subsection{Transport properties of a clean screw dislocation}\label{section3.1}

\begin{figure*}
\includegraphics[scale=0.2]{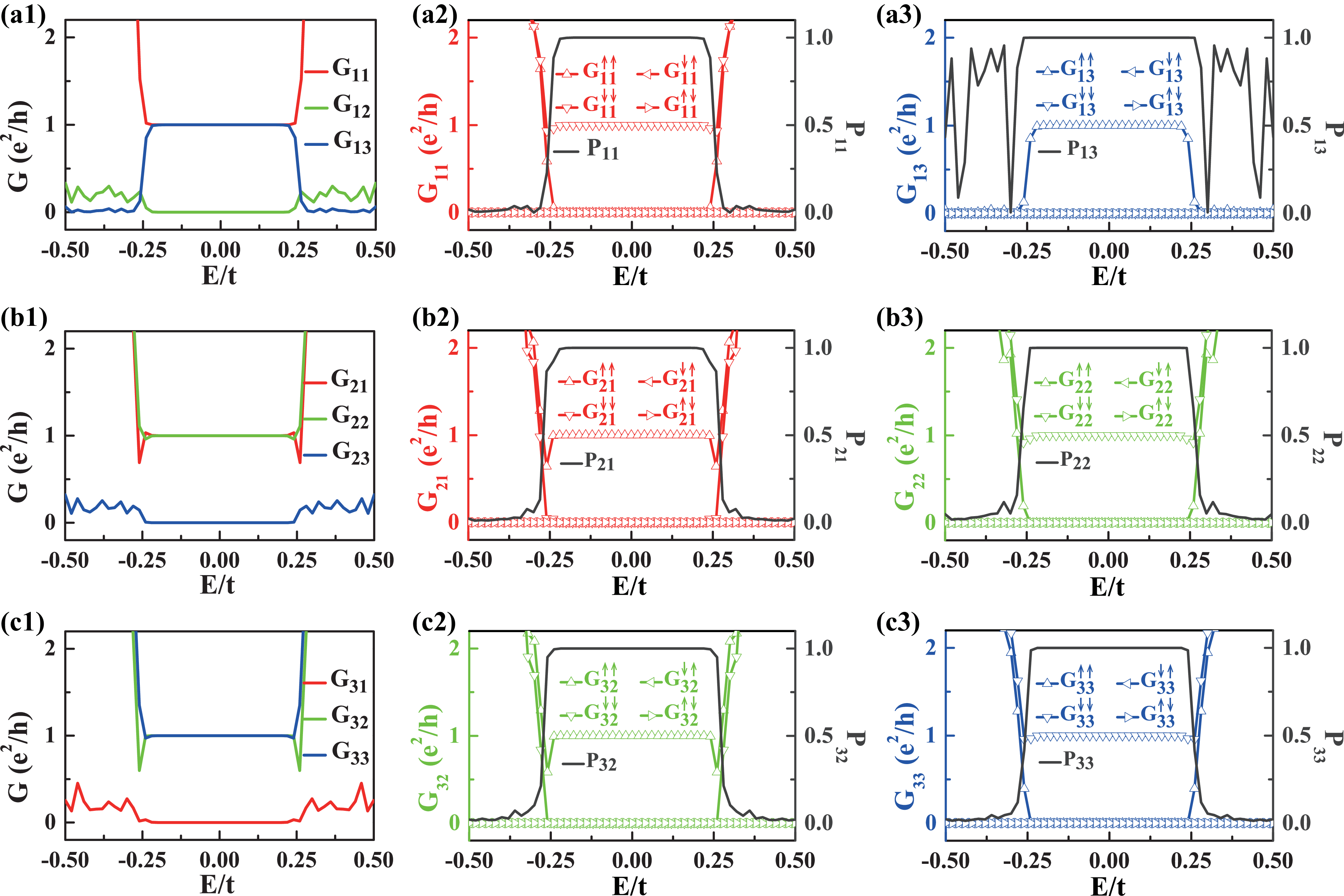}
\caption{\label{fig:2} The conductance of a multi-terminal layer-spin filter. (a1), (b1) and (c1) are the conductance form the source layer S1, S2, and S3, respectively. All the drain layers are able to output carriers. (a2)-(a3), (b2)-(b3) and (c2)-(c3) are their corresponding spin conductance and polarization. The size is $N=25$ and $M=50$.}
\end{figure*}

In Fig. \ref{fig:2}, we present the main transport results, including the linear conductance, spin polarized conductance, and corresponding spin polarization of the current flow for three cases: terminal S1, S2, or S3 is used as the source layer, respectively.

\begin{figure}[b]
\includegraphics[scale=0.21]{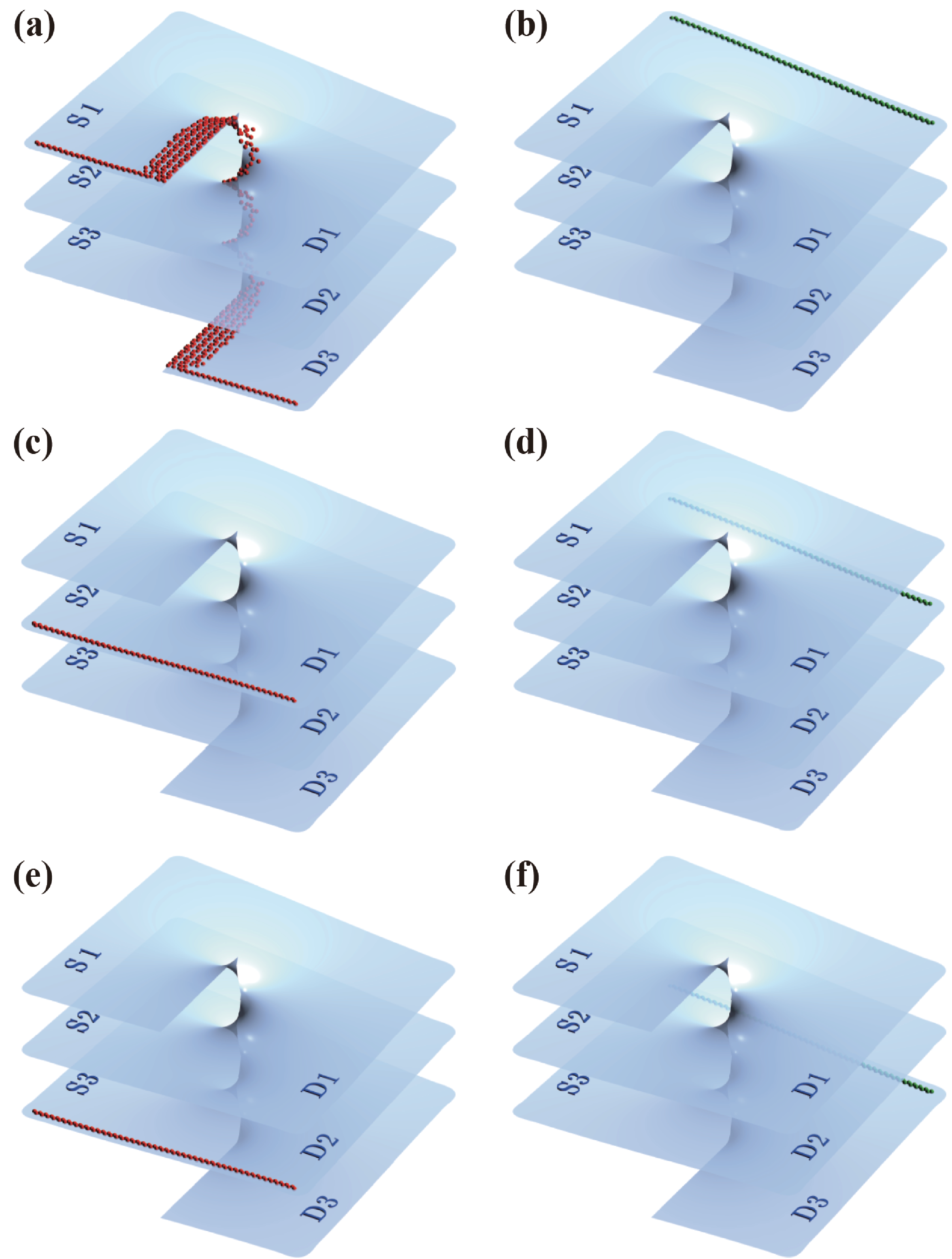}
\caption{\label{fig:3} The spatial distribution of spin polarized current of the multi-terminal device with a screw dislocation. The input of (a) and (b) is source layer S1; The input of (c) and (d) is S2; (e) and (f) are from source layer S3. All the drain layers are capable of outputting. The parameters are the same with Fig. \ref{fig:2}.
}
\end{figure}

In Fig.~\ref{fig:2}(a1)--\ref{fig:2}(a3), current is emitted from the source layer S1. As shown in Fig.~\ref{fig:2}(a1), within the topological band gap, only \( G_{11} \) and \( G_{13} \) exhibit quantized conductance of \(e^2/h \), while \( G_{12} \) is zero. This behavior arises because S1 is directly connected to D1 via the top layer. Meanwhile, S1 is also linked to the D3 via the screw dislocation, as shown in Fig.~\ref{fig:3}(a) and \ref{fig:3}(b).
Consequently, the spin-polarized current flows along the front and back boundaries toward the drain layers D3 and D1, respectively. Outside the topological band gap, the value of \( G_{11} \) increases rapidly, \( G_{12} \) remains small, and \( G_{13} \) is zero. In this energy region, the bulk states dominate the transport. The results suggest that only the topological state can propagate through the screw dislocation, whereas bulk states are unable to participate in the transmission along this path. The interlayer channel along the screw dislocation functions as an elevator, allowing topological spin polarized current to propagate between the top and bottom layers. The majority of carriers in the bulk states flow from S1 to D1, with a small fraction scattering into the second layer and exiting through the layer D2.

To support the above explanation, we also computed the spin-resolved transmission and spin polarization from source layer S1 to drain layer D1 and D3, as shown in Fig.~\ref{fig:2}(a2) and \ref{fig:2}(a3). Within the topological band gap, \( G_{11}^{\downarrow\downarrow} \) and \( G_{13}^{\uparrow\uparrow} \) equal to $e^2/h$, which means that within this energy range, the spin-down current from S1 flows into D1, and the spin-up current flows into D3. In other words, the spin polarization $P_{11}$ and $P_{13}$ are equal to $1$ in this energy region. When the Fermi energy $E$ lies outside the topological band gap, bulk states participate in the transport, and \( G_{11} \) increases rapidly. In this case, the bulk current flowing from S1 to D1 dominates, and the spin polarization $P_{11}$ rapidly decreases to $0$. In this energy range, the transmission from the top layer to the bottom layer is nearly negligible: \( G_{13} = 0 \). This result indicates that, when S1 is chosen as the input source layer, the topological spin-down current is output at drain layer D1, while the topological spin-up current is output at drain layer D3.
To provide an intuitive visualization of the spatial transport of spin-polarized QSH edge states, the corresponding local distributions of spin-polarized current are also presented in Fig.~\ref{fig:3}. The results are calculated from Eq. \ref{current}. The transmission processes are clearly presented by Fig. \ref{fig:3}(a) and \ref{fig:3}(b).

Fig.~\ref{fig:2}(b1)--\ref{fig:2}(b3) corresponds to the case that current is emitted from the source layer S2. As shown in Fig.~\ref{fig:2}(b1), within the energy gap \( G_{21} \) and \( G_{22} \) exhibit quantized conductance and \( G_{23}=0 \).
This behavior can be directly explained by the spin polarized current flow shown in Fig. ~\ref{fig:3}(c) and \ref{fig:3}(d).
Since S2 is connected to D1 and D2 via the front edge and back edge respectively, spin-up (down) current flows into drain layer D1 (D2). It is also supported by the spin resolved transmission results.
In Fig.~\ref{fig:2}(b2) and \ref{fig:2}(b3), we can see that only \( G_{21}^{\uparrow\uparrow} \) and \( G_{22}^{\downarrow\downarrow} \) have non-zero values within the gap. The spin-up current flows from source layer S2 into drain layer D1, while the spin-down current flows from source layer S2 into drain layer D2. Besides, the current is totally spin polarized: $P_{21}=P_{22}=1$. Outside the topological band gap, bulk states participate in the transport, and \( G_{11} \) and \( G_{12} \) both increase rapidly. Bulk states are free to propagate either to D1 along the front half or to D2 along the back half. And the spin polarization $P_{21}$ and $P_{22}$ are close to $0$.
This means that if S2 is chosen as the input source layer, the topological spin-up state flows into drain layer D1, while the topological spin-down state flows into drain layer D2.

The case with S3 as the input source layer was also examined in Fig.~\ref{fig:2}(c1)--\ref{fig:2}(c3) and~\ref{fig:3}(e)--\ref{fig:3}(f). Within the topological band gap, \( G_{32} \) and \( G_{33} \) are both \(e^2/h \) as shown in Fig. ~\ref{fig:2}(c1). In Fig.~\ref{fig:2}(c2), \( G_{32}^{\uparrow\uparrow} =e^2/h \) and \( P_{32} = 1 \). In Fig.~\ref{fig:2}(c3), \( G_{33}^{\downarrow\downarrow} = e^2/h \) and \( P_{33} = 1 \). That is, the spin-down current propagates along the back boundary to drain layer D3, while the spin-up current propagates along the front boundary to drain layer D2, as depicted in Fig.~\ref{fig:3}(e) and \ref{fig:3}(f).

\begin{table}
\centering
\begin{tabular}{c c c c}
\hline
~         & Drain 1 & Drain 2 & Drain 3 \\
\hline
Source 1  & $\downarrow$ & ~       & $\uparrow$ \\
Source 2  & $\uparrow$   & $\downarrow$ & ~ \\
Source 3  & ~       & $\uparrow$   & $\downarrow$ \\
\hline
\end{tabular}
\caption{\label{tab:table1}  The layer-spin filtering rules of a single screw dislocations. }
\end{table}

\begin{figure}[b]
\includegraphics[scale=0.26]{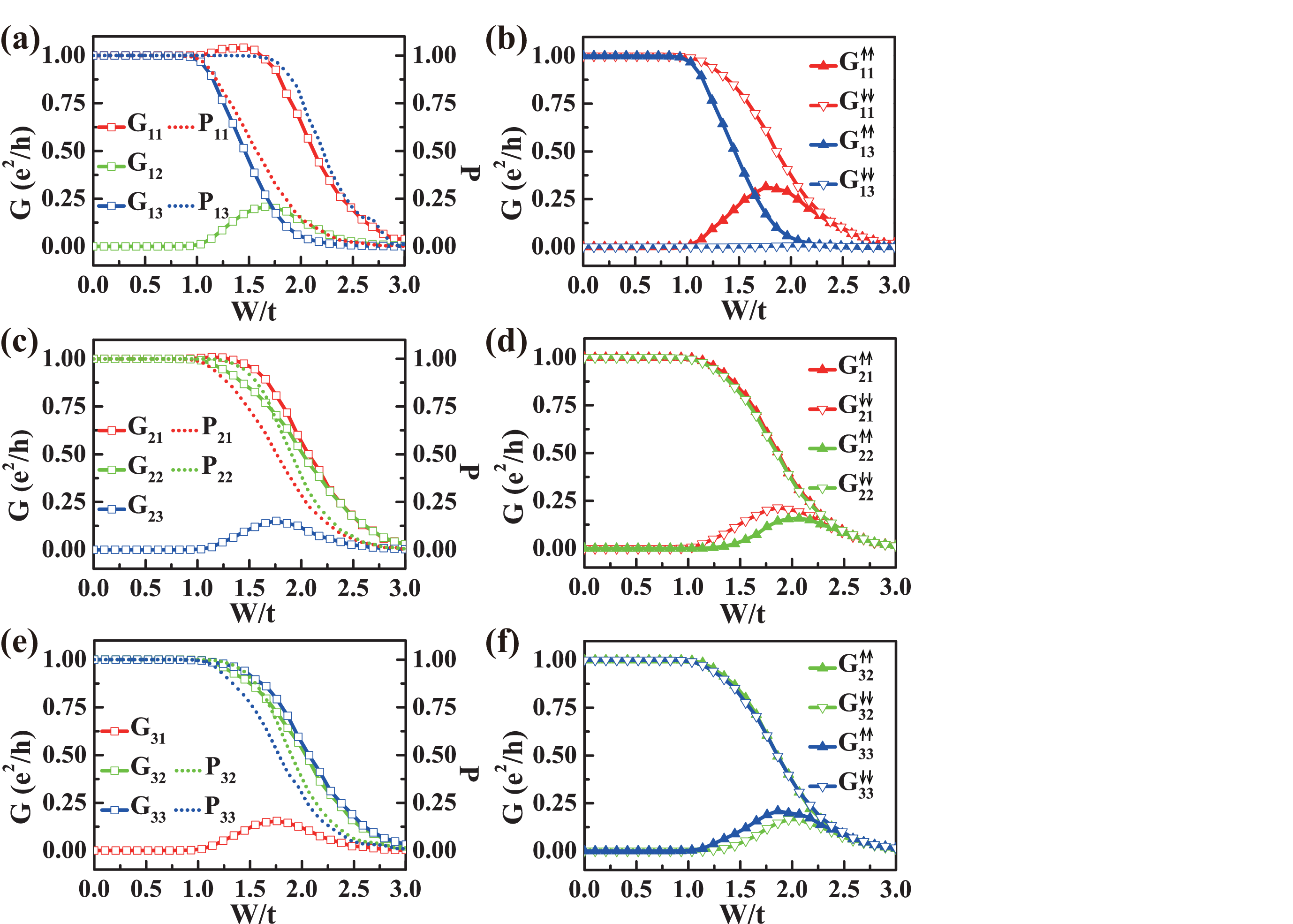}
\caption{\label{fig:4} The conductance and spin polarization v.s. Anderson disorder strength. (a) and (b) for S1 input; (c) and (d) for S2 input; (e) and (f) for S3 input. The Fermi energy is set as $E=0.1t$. Other parameters are the same with Fig.~\ref{fig:2}.}
\end{figure}

Based on the above analysis, we can derive the following rule of layer-spin selection. In the Kane-Mele model, the nanoribbon hosts QSH edge states, where different spin states propagate in opposite directions. When the QSH edge states at the source pass through the multilayer screw dislocation structure, fully spin-polarized currents can be generated at specific drain layers. The structure contains two types of transport channels: one runs along the boundaries, and the other passes through the screw dislocation. In the structure shown in Fig.~\ref{fig:3}(a), only the spin-up carriers, starting from the terminal S1, will pass through the screw dislocation and flow to the bottom drain layer D3. For the spin-up states that starts from the source layer S$n$ ($n \neq 1$), they will end at the drain layer D$m$ ($m=n-1$). On the other hand, for the spin-down carriers that start from the source layer S$n$, they will flow to the drain layer D$n$. This rule is detailed in Table~\ref{tab:table1}.
While pioneering studies have demonstrated topological layer filtering in specific systems\cite{AGao,DZhai,WBDai}, the synergistic control of spin-resolved transport with layer selectivity remains unexplored--a gap our work now bridges.

\begin{figure}[b]
\includegraphics[scale=0.26]{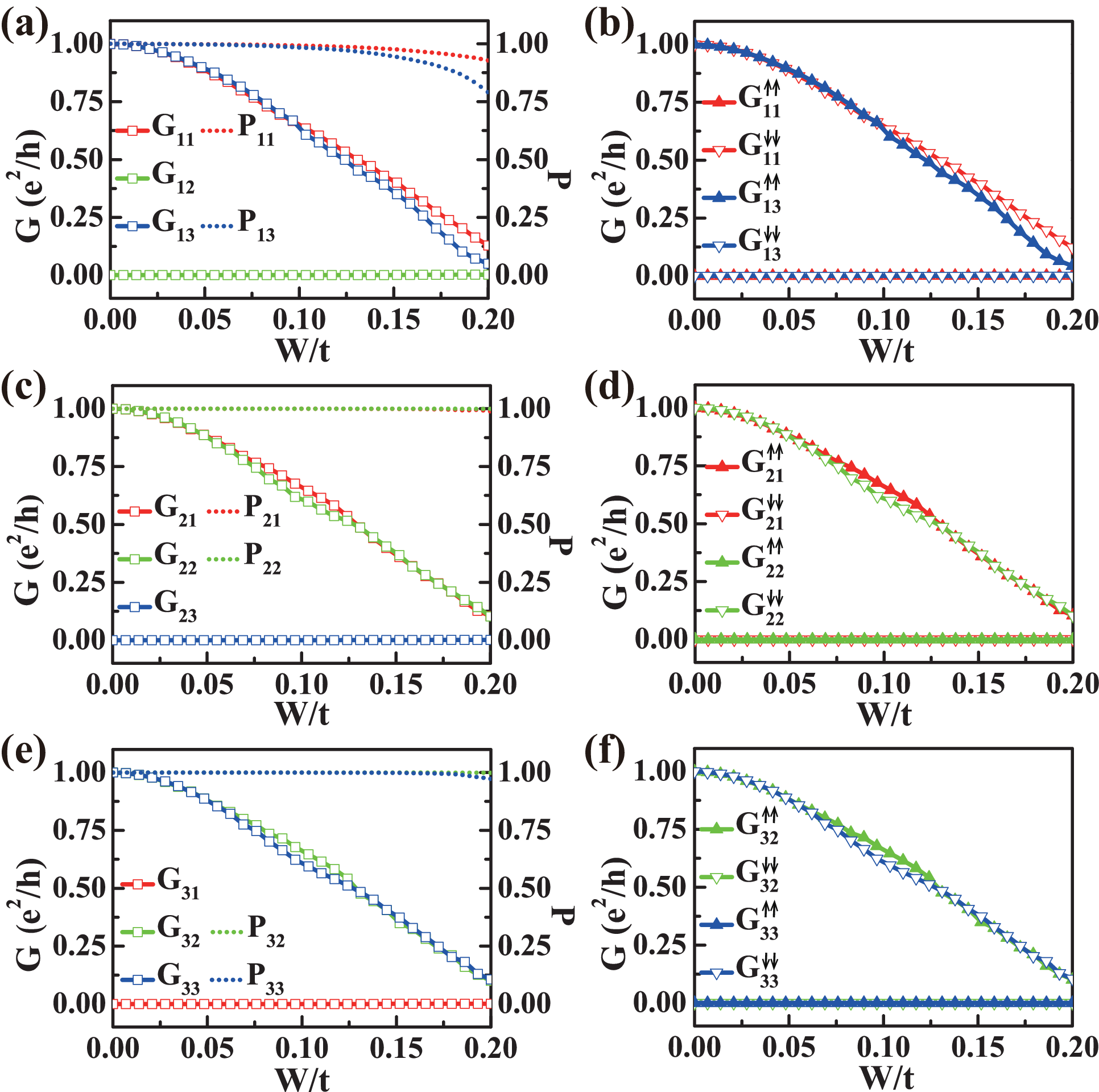}
\caption{\label{fig:5} The conductance and spin polarization v.s. magnetic disorder. (a) and (b) for S1 input; (c) and (d) for S2 input; (e) and (f) for S3 input. The Fermi energy is set as $E=0.1t$. Other parameters are the same with Fig.~\ref{fig:2}.}
\end{figure}

By selecting specific source layers, the spin orientation of the output current at each drain layer can be controlled. For example: i) Selecting S1 fixes D1 to spin-down and D3 to spin-up, while no current flows into D2. ii) Selecting S2 fixes D1 to spin-up and D2 to spin-down, with no current to D3. iii) Selecting S3 fixes D2 to spin-up and D3 to spin-down, with no current to D1. This demonstrates that the spin polarization at a given drain layer can be altered by changing the source layer, enabling precise control over spin-polarized currents, providing a versatile layer-spin filter platform.

\subsection{Transport properties of a disordered screw dislocation }\label{section3.2}

Traditional QSH states are robust against Anderson disorder and sensitive to magnetic disorders. Thus, we wonder how the present layer-spin filter works under disorder. In the above discussion, the spin polarized current emitted from each source layer can propagate via two distinct sets of transmission paths, as shown in Fig.~\ref{fig:3}. One set follows the path of the screw dislocation, while the other runs along the outer boundaries. The first kind has a longer route than the latter one.

Here, we focus on the effect of disorder on the transport of QSH states, thus the Fermi level is set to \(E = 0.1t\), which lies within the topological band gap. To ensure data accuracy, all conductance data were averaged over $3000$ calculations.
In Fig.~\ref{fig:4}, the effects of Anderson disorder on the transport properties of the QSH edge state are displayed for three types of input, S1, S2, and S3. The case for input S1 is special. It has two distinct channels: one from the outer boundary and the other via the screw dislocation. The cases for input S2 and S3 are similar, both have two outer boundary channels.

In general, the device operates robustly. As shown in Fig.~\ref{fig:4}(a), even when the disorder strength is relatively high (e.g., $W=t$), the quantized quantities (e.g. $G_{11}$ and $G_{13}$) of the clean case remain well preserved, and the zero value ($G_{12}$) of the clean case remain zero.
However, as $W$ increases further, several characteristics appear: i) $G_{11}$ first increases from $e^2/h$ to a large value and then decreases to zero; ii) $G_{12}$ first increases from zero and decays when $W$ is further increased; iii) $G_{13}$ decreases monotonically. The results signify that around $W=1.5t$, the system enters into the diffusive regime. In this scenario, the current flowing into D1 should contain spin-down states (since $G>e^2/h$). Then we wonder, where scattering occurs and which states contribute to $G_{12}$. To investigate the details, we calculate the spin components of the conductances in Fig.~\ref{fig:4}(b). Since spin is a good quantum number under Anderson disorder, we do not consider spin-flip during the transport process in Fig.~\ref{fig:4}, i.e., $G^{\uparrow\downarrow}=G^{\downarrow\uparrow}=0$.

The results demonstrate that, in the diffusive regime, the output of drain layer D1 does contain both spin-down current and a significant part of spin-up component. Meanwhile, $G_{12}$ is contributed by spin-up states from source layer S1 (not shown). Thus we speculate that the physics picture should be depicted as follows. Firstly, the transmission of spin-down carriers from S1 to D1 is rather straightforward. They simply flow along the back edge of the top layer. Under disorder, a fraction of them are backscattered into S1 and the proportion flowing into D1 decreases monotonically as $W$ is enhanced.
Secondly, let us track the spin-up carriers flowing out from S1.
The carriers from S1 move to the top end of screw dislocation and are partially scattered. A fraction of them are scattered into D1 via the front edge of top layer and the remaining part continues to descend in a spiral along the screw dislocation. On the next layer, again the spin-up carriers are scattered and part of them flow into D2 via the front edge. Finally the rest follow the path shown in Fig.~\ref{fig:3}(a) into D3.
The above discussion explains all the three characteristics we proposed in the previous paragraph. In brief, in the diffusive regime the current flowing into D1 contains a spin-down component and the current flowing into D3 consists of a gradually diminishing spin-up component. It thus explains why in Fig.~\ref{fig:4}(a),
\( P_{11} \) decreases to small values even when $G_{11}$ is still larger and \( P_{13} \) is still near completely polarized when $G_{13}$ is small. Figure ~\ref{fig:4}(b) also has a notable feature worth discussing, which is that $G^{\uparrow\uparrow}_{13}$ decreases faster than $G^{\downarrow\downarrow}_{11}$. It is due to the fact that spin-up carriers experience a longer route from the source to the drain than spin-down carriers. As a result, they undergo stronger scattering.

For cases with input from source layer S2 and S3, the results are displayed in Fig.~\ref{fig:4}(c)-\ref{fig:4}(f). Since the paths for both spin-up and spin-down states are identical, \( G_{21} \) and \( G_{22} \) (as well as \( G_{32} \) and \( G_{33} \)) exhibit nearly identical transport properties under Anderson disorder. Taking input source layer S2 as an example [see Fig.~\ref{fig:4}(c) and \ref{fig:4}(d)], in the diffusive regime ($\sim W \in [t,2.5t]$), spin-up carriers from S2 can transmit into D1 or be scattered to D2 via the lower layer.
The spin-down carriers, in addition to flowing directly to D2, can also be scattered to D1 by climbing up one layer or descend one layer to D3. It explains the nonzero of $G_{23}$.
Since $G^{\downarrow\downarrow}_{21}$ and $G^{\uparrow\uparrow}_{22}$ are rather close to each other, $P_{21}$ and $P_{22}$ are similar as well. Finally, for very large $W$, the system enters into the insulating area and all transmission components decay to zero.

\begin{figure}[b]
	\includegraphics[scale=0.16]{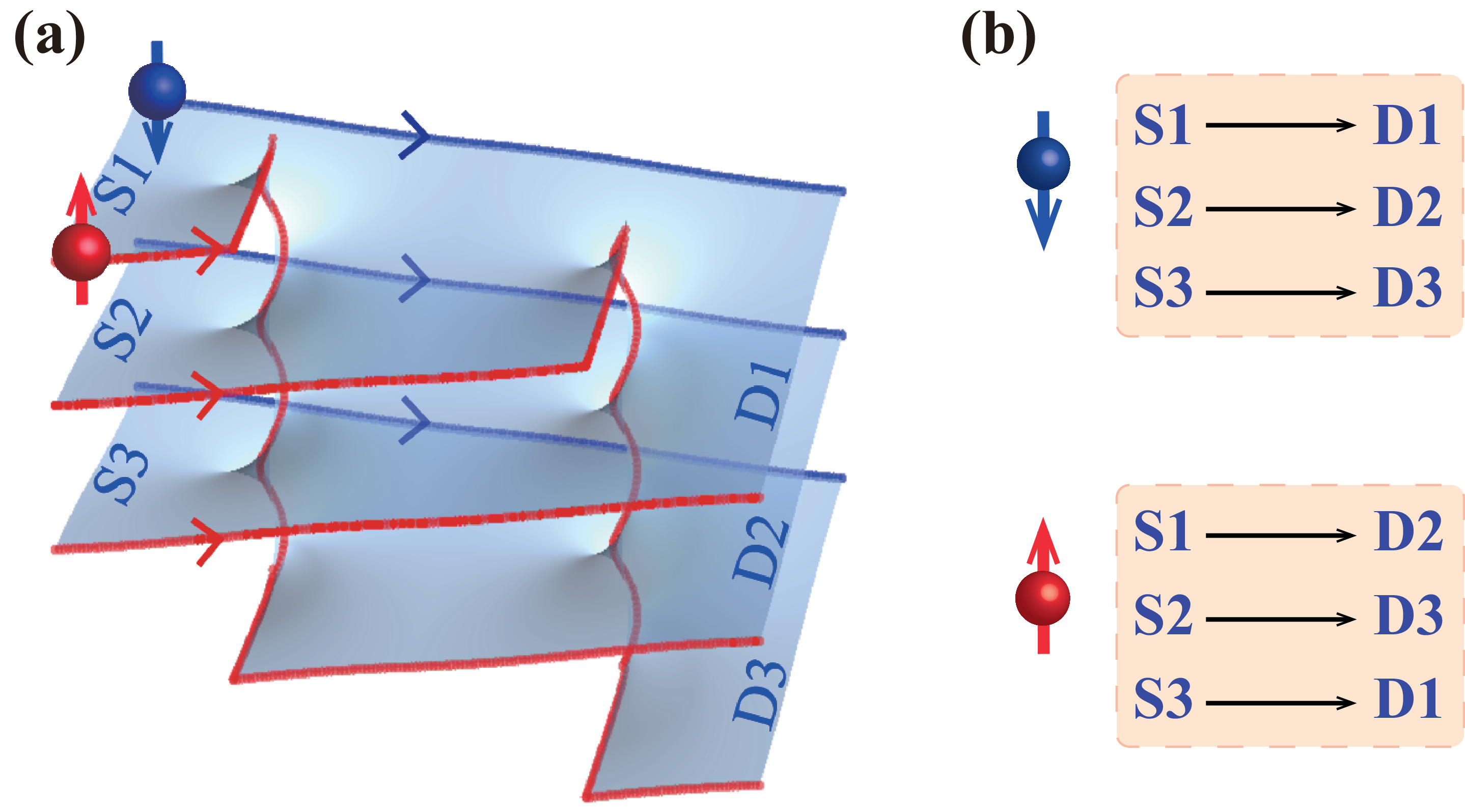}
	\caption{\label{fig:6} (a) Schematic of the device with two identical screw dislocations. The drain layers remain connected, and the source layers S1, S2, and S3 are selectively activated. The topological channels associated with the QSH edge states are represented by blue (spin-down) and red (spin-up) lines. (b) Layer-spin filtering of device in (a) . }
\end{figure}

\begin{figure*}
	\includegraphics[scale=0.2]{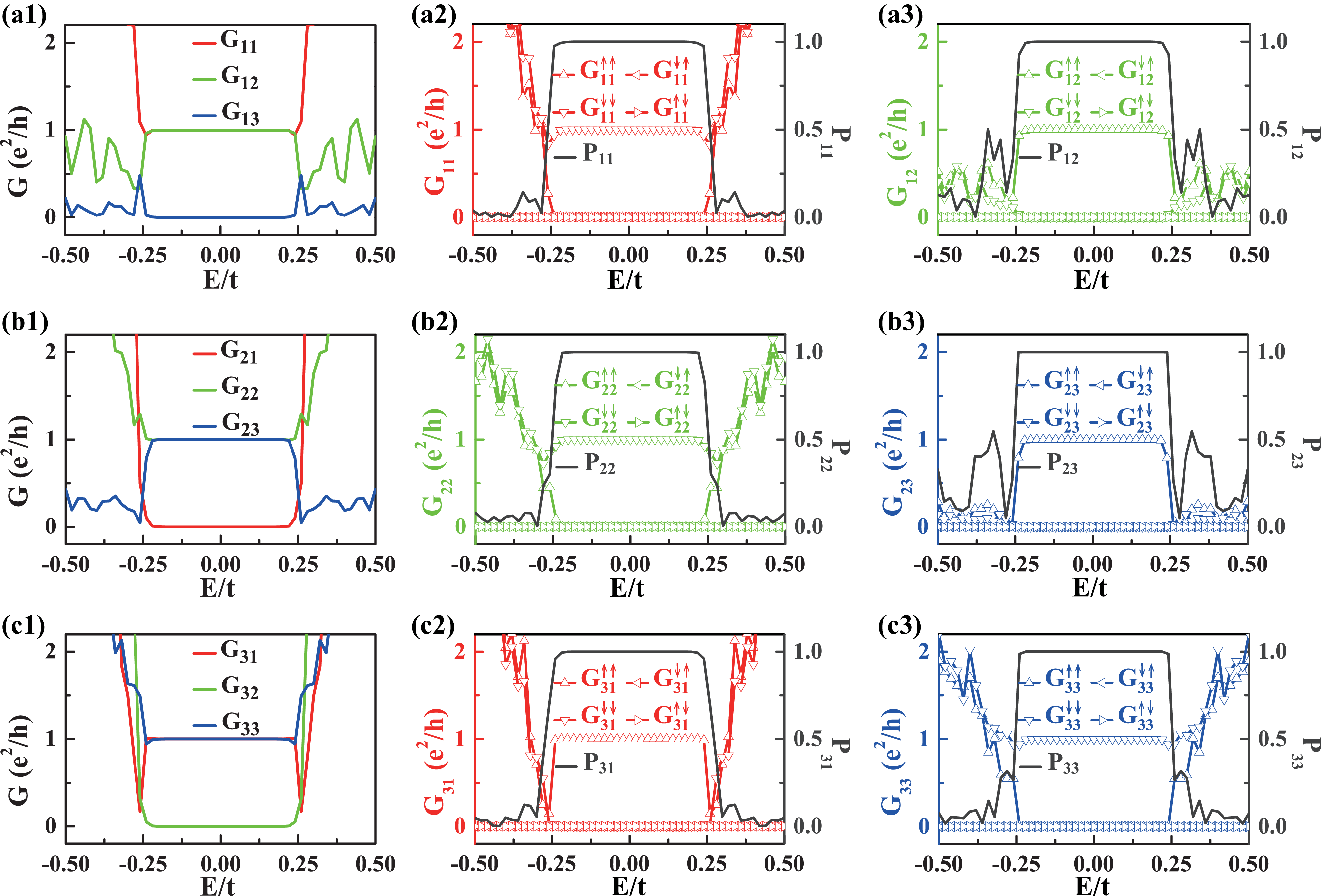}
	\caption{\label{fig:7} The conductance of the device in Fig.~\ref{fig:6}(a). (a1), (b1) and (c1) are the conductance form the source layer S1, S2, and S3, respectively. All the drain layers are connected, enabling output charge carriers. (a2)-(a3), (b2)-(b3) and (c2)-(c3) are their corresponding spin-resolved conductance and spin-polarization. The size of the device is $N=25$ and $M=100$.}
\end{figure*}
  
It is well known that QSH states are fragile to magnetic disorder. Thus, we investigate how the layer-spin device works under magnetic disorder. The results are collected in Fig.~\ref{fig:5}. For all three cases (input S1, S2, and S3), the quantized transmission coefficients decay to zero monotonically when $W$ is increased (even to a small value $W=0.2t$). The spin components of the linear conductance in Fig.~\ref{fig:5}(b), \ref{fig:5}(d), and \ref{fig:5}(f) show same features: the non-zero ones decay from $e^2/h$ to zero, while the ones equal to zero never increase. Meanwhile, the spin-polarization values for almost all cases stay at $P=1$. The results indicate that the transmission coefficients of QSH states in the present device is sensitive to magnetic disorder. Spin flip induced backscattering occurs at relatively small $W$, well before edge states have any chance to be scattered from one edge to the other while preserving their spin index, as the latter requires stronger disorder. In this way, the fully spin-polarized current flows to different layers, and the system retains good layer-spin properties despite being in magnetic disorder.

\subsection{Transport properties of two clean screw dislocations }\label{section3.3}
Having characterized the layer-spin transport properties in devices with individual screw dislocations, we now examine the corresponding behavior in systems containing a pair of screw dislocations.
Figure~\ref{fig:6} shows a composite structure obtained by joining two Fig.~\ref{fig:1}(b) configurations. In Fig.~\ref{fig:6}(a), all available channels are shown, where blue curves represent spin-down states and red curves denote spin-up states.
Following the layer-spin filtering rule in Table~\ref{tab:table1}, we summarize the layer-spin filtering rule for two screw dislocations in Fig.~\ref{fig:6}(b): The route of spin-down carriers is from S$m$ to D$m$ and the route of spin-up carriers is from S$m$ to $\rightarrow$ D$n$ ($n=m-2$ or $m+1$). Compared to the layer-spin filtering rules governing individual screw dislocations in section \ref{section3.1}, the rules for the present two-dislocation system exhibit an additional cyclic permutation symmetry.

To validate this conjecture, the transport properties were calculated, as presented in Fig.~\ref{fig:7}. As a specific example, we configure S1 as the source electrode, with the corresponding results presented in Fig.~\ref{fig:7}(a1)-\ref{fig:7}(a3). Within the topological band gap $[-0.26t, 0.26t]$, only $G_{11}$ and $G_{12}$ have quantized conductance, exclusively arising from $G_{11}^{\downarrow\downarrow}$ and $G_{12}^{\uparrow\uparrow}$, respectively. This indicates that spin-down current exits via terminal D1, while spin-up current exits via terminal D2. In this range, both $P_{11}$ and $P_{12}$ are equal to 1. In summary, for devices incorporating multiple screw dislocations, the underlying mechanism effectively provides successive selection in both layer and spin degrees of freedom. This architecture simultaneously maintains quantized spin conductance and achieves high spin polarization, thereby offering a novel approach to layer-spin filtering.

\section{CONCLUSION AND PROSPECTS}
In summary, we reported that a screw dislocation structure based on Kane-Mele graphene can serve as a perfect layer-spin filter. The QSH edge states are capable of propagating along the boundary to neighboring layers or traversing from the top to the bottom layer through screw dislocations. By selecting different layers for the source, the current and its spin index in specific drains layer are determined accordingly. Both the topological spin currents along the outer boundaries and through the screw dislocation remain robust under Anderson disorder. Under magnetic disorder, spin-flip induced backscattering occurs. While exhibiting significant transmission attenuation in all source layers, the system maintains robust layer-spin filtering functionality.
Furthermore, the layer-spin selecting transport characteristics for a device incorporating two screw dislocations have been examined, revealing findings that closely mirror those observed in a single screw dislocation scenario.

In the discussion of a device with two screw dislocations in section \ref{section3.3}, it is interesting to investigate the transport interference of two separate screw dislocations. The realization of quantum interference here requires two conditions: (1) Two paths propagating through the screw dislocations should share the same source and drain to form a complete loop. (2) Back scattering between the screw dislocation state and the edge state. Let's elaborate on the second point. In the absence of backscattering, within the energy gap, the conductance should always remain a constant value (specifically, $2e^2/h$, which is contributed by one spin - up state and one spin - down state), and there will be no interference. There are two ways to introduce backscattering. For example, we can take disorder into account or reduce the spatial distance between the two states of the same spin. The interference effects of screw dislocations provide critical evidence for their existence and conduction properties, establishing a powerful methodology for studying 1D channels in 3D crystalline materials. We anticipate this approach will inspire further advances in the field.

\section*{ACKNOWLEDGMENTS}

\end{document}